\documentclass[11pt]{article}
\usepackage{amsmath,amsfonts,latexsym,graphicx,amssymb}
\date{}
\def\la{\langle\,}
\def\r{\,\rangle}

\newcommand{\bbR}{\mathbb{R}}
\newcommand{\bbC}{\mathbb{C}}
\newcommand{\bbN}{\mathbb{N}}
\def\oper{{\mathchoice{\rm 1\mskip-4mu l}{\rm 1\mskip-4mu l}%
{\rm 1\mskip-4.5mu l}{\rm 1\mskip-5mu l}}}
\def\con{{}_{\_\rule{-1pt}{0pt}\_}
\rule{-2pt}{0pt}\raise1.5pt\hbox{$\mid$}\hspace{2pt}}

\newtheorem{theorem}{Theorem}
\newtheorem{proposition}{Proposition}
\newtheorem{corollary}{Corollary}
\title{\bf Quantum Mechanics of Damped Systems}
\author{Dariusz Chru\'sci\'nski\footnote{On the leave from
 Institute of Physics, Nicolaus Copernicus University,
 ul. Grudzi\c{a}dzka 5/7, 87-100 Toru\'n, Poland} \\
Department of Mathematics and Statistics \\
University of Calgary \\ Calgary, Alberta, Canada}

\begin{document}

\maketitle

\begin{abstract}

We show that the quantization of a simple  damped system leads to
a self-adjoint Hamiltonian with a family of complex generalized
eigenvalues. It turns out that they correspond to the poles of
energy eigenvectors when continued to the complex energy plane.
Therefore, the corresponding generalized eigenvectors may be
interpreted as resonant states. We show that resonant states are
responsible for the irreversible quantum dynamics of our simple
model.

\end{abstract}

\vspace{.7cm}


\vspace{.3cm}


\newpage

\section{Introduction}
\setcounter{equation}{0}

\numberwithin{equation}{section}

Standard textbooks on quantum mechanics investigate mainly the
Hamiltonian system, i.e. by a quantum system one usually means a
Hilbert space $\cal H$ which describes physical quantum states and
a self-adjoint operator (Hamiltonian) in $\cal H$ which governs
dynamics of the system. However, most of the classical systems
 are not Hamiltonian and the quantum mechanics
of such systems is poorly understood. In the present paper we are
going to investigate one of the simplest non-Hamiltonian system
corresponding to a damped motion in one dimension:
\begin{eqnarray} \label{damp} \dot{x} = -\gamma x \ , \end{eqnarray} where
$x \in \bbR$, and $\gamma > 0$ stands for the damping constant.
Classically, the damping behavior is described by the exponential
law \begin{eqnarray} x(t) = e^{-\gamma t} x\ . \end{eqnarray} As
is well known \cite{Khalfin} (see also \cite{Hegerfeldt}), within
the standard Hilbert space formulation of quantum mechanics there
is no room for such a behaviour on a quantum level. Therefore, in
order to deal with this problem, we shall use the rigged Hilbert
space approach to quantum mechanics which generalizes the standard
Hilbert space version \cite{Bohm,Bohm-Gadella,Prigogine}. A rigged
Hilbert space (or a Gelfand triplet) is a collection of spaces
\cite{RHS1,RHS2}: \begin{eqnarray} \Phi \subset {\cal H} \subset
\Phi'\ , \end{eqnarray} where $\cal H$ is a Hilbert space, $\Phi$
is a dense nuclear subspace of $\cal H$, and $\Phi'$ denotes its
dual, i.e. the space of continuous functionals on $\Phi$ (see
section~\ref{RHS} for a brief review).

The quantization of our simple model (\ref{damp}) leads to a
self-adjoint  Hamiltonian $\widehat{H}$ in ${\cal H} = L^2(\bbR)$.
Interestingly, $\widehat{H}$ being  self-adjoint, gives rise to
the family of generalized complex eigenvalues. Clearly, these
eigenvalues are not elements of the spectrum $\sigma(\widehat{H})
=(-\infty,\infty)$. The corresponding  eigenvectors do not belong
to $L^2(\bbR)$ but to $\Phi'$ for an appropriately chosen $\Phi$.
We show that these complex eigenvalues  have many remarkable
properties analogous to the point spectrum of a self-adjoint
operator.  In particular, they give rise to the spectral
decomposition of $\widehat{H}$. Moreover, they are closely related
to the continuous spectrum of $\widehat{H}$. It turns out that
they correspond to the poles of the  energy eigenvectors $\psi^E$
when continued to the complex energy plane \cite{Gorini}.
Physicists usually called the corresponding eigenvectors resonant
states \cite{Bohm,RES-1,RES-2} (see also \cite{Reed}). It is
widely believed that resonant states are responsible for the
irreversible dynamics of physical systems (see e.g. recent
collection of papers \cite{Bohm}). Indeed, it is true in our
simple model. To see this we
 construct two Gelfand triples:
\begin{eqnarray} \Phi_\pm \subset L^2(\bbR) \subset \Phi_\pm'\ , \end{eqnarray} such
that $\Phi_+ \cap \Phi_- = \{\emptyset\}$. Obviously, the time
evolution is perfectly reversible  when considered on $L^2(\bbR)$.
It is given by the 1-parameter group of unitary transformations
$U(t)=e^{-i\widehat{H}t}$. However,
 when restricted to $\Phi_\pm$, it defines only two
semigroups: $U(t\geq 0)$ on $\Phi_-$, and $U(t\leq 0)$ on
$\Phi_+$. Therefore, the evolution on $\Phi_\pm$ is irreversible.
This irreversibility is caused by quantum damping, or,
equivalently, by the presence of resonances.

\section{Rigged Hilbert space}
\label{RHS}

Consider a rigged Hilbert space, i.e. the following  collection
(Gelfand triplet): \begin{eqnarray} \Phi \subset {\cal H} \subset
\Phi'\ , \end{eqnarray} where $\cal H$ is a Hilbert space with the
standard norm topology $\tau_{\cal H}$, $\Phi$ is a topological
vector space with a topology, $\tau_{\Phi}$, stronger than
$\tau_{\cal H}$, and $\Phi'$ is the dual space of continuous
linear functionals on $\Phi$ \cite{RHS1,RHS2}. We denote the
action of $\Phi'$ on $\Phi$ using Dirac notation, i.e. for any
$\phi \in \Phi$ and $F\in \Phi'$ \begin{eqnarray} \la \phi | F\r
:= F(\phi)\ . \end{eqnarray} Any self-adjoint operator
$\widehat{A}$ in $\cal H$ may be extended to an operator on
$\Phi'$:
\begin{eqnarray} \widehat{A} : \Phi' \rightarrow \Phi'\ , \end{eqnarray} by
\begin{eqnarray} \la \phi | \widehat{A} F\r := \la \widehat{A}\phi
| F\r \ . \end{eqnarray} Now, if for any $\phi \in \Phi$
\begin{eqnarray} \la \phi|\widehat{A} F_\lambda\r = \lambda \la
\phi|F_\lambda \r \ , \end{eqnarray} then $F_\lambda \in \Phi'$ is
called a generalized eigenvector corresponding to a generalized
eigenvalue $\lambda$. Omitting $\phi$ one simply writes:
\begin{eqnarray} \widehat{A}|F_\lambda\r = \lambda |F_\lambda\r \
. \end{eqnarray} Note, that a generalized eigenvalue $\lambda$ may
be complex. Now, if the spectrum of $\widehat{A}$ \begin{eqnarray}
\sigma(\widehat{A}) = \sigma_p(\widehat{A}) \cup
\sigma_c(\widehat{A}) \ \subset \ \bbR\ , \end{eqnarray} with
$\sigma_p(\widehat{A}) = \{ \lambda_1, \lambda_2, \ldots \}$, then
the Gelfand-Maurin theorem \cite{RHS1,RHS2} implies the following
spectral decompositions: \begin{eqnarray} \oper_{\Phi} = \sum_n
|F_n\r \la F_n| + \int_{\sigma_c(\widehat{A})} d\lambda \,
|F_\lambda\r\la F_\lambda|\ , \end{eqnarray} and of $\widehat{A}$
itself:
\begin{eqnarray} \widehat{A} = \sum_n \lambda_n |F_n\r \la F_n| +
\int_{\sigma_c(\widehat{A})} d\lambda \, \lambda |F_\lambda\r\la
F_\lambda|\ . \end{eqnarray} This way the rigged Hilbert space
approach fully justifies the standard Dirac notation.

The choice of $\Phi$ depends on the particular problem one deals
with. In the present paper we shall consider the following
functional spaces: $\cal D$ -- the space of $C^\infty(\bbR)$
functions with compact supports equipped with the convex Schwartz
topology \cite{Yosida}, $\cal S$ -- the space of $C^\infty(\bbR)$
functions vanishing at infinity faster than any polynomial
\cite{Yosida}. Moreover, let us define \begin{eqnarray} {\cal Z}
:= \{ F[\phi]\, |\, \phi \in {\cal D}\}\ , \end{eqnarray} where
$F[\phi]$ denotes the Fourier transform of $\phi$. It turns out
\cite{Gelfand-S} that $\cal Z$ is isomorphic to the space of
entire functions of fast decrease along $\bbR$. More precisely,
let \begin{eqnarray} F_L[\phi](z) :=
 \frac{1}{\sqrt{2\pi}} \int_{-\infty}^\infty e^{izx}\phi(x)\, dx\ ,
\end{eqnarray}
be the Fourier-Laplace transform of
 $\phi \in {\cal D}$. One proves \cite{Yosida,Gelfand-S} the following
\begin{theorem}[Paley-Wiener-Schwartz]
Let $a>0$. An entire  function $U(z)$ is a Fourier-Laplace
transform of a function $u\in {\cal D}$ with support
\[  {\rm supp}(u)=\{x\in \bbR\,|\, |x|\leq a\}\ , \]
 if and only if
\begin{eqnarray}
|z|^n|U(z)| \leq C_ne^{a|{\rm Im}z|} \  ,\ \ \ \ \ \ \
n=1,2,\ldots\ .   \nonumber
\end{eqnarray}
\end{theorem}
Now, for $z=x \in \bbR$, i.e. ${\rm Im}\, z=0$, $F_L[\phi] =
F[\phi]$, and the above theorem implies \begin{eqnarray} |x|^n
|F[\phi](x)| \leq C_n \  ,\ \ \ \ \ \ \ n=1,2,\ldots\ .
\end{eqnarray} Clearly, ${\cal Z} \cap {\cal D} = \{\emptyset\}$.
Moreover, one has \begin{eqnarray} {\cal D} \subset {\cal S}
\subset L^2(\bbR)\ , \end{eqnarray} and \begin{eqnarray} {\cal Z}
\subset {\cal S} \subset L^2(\bbR)\ , \end{eqnarray} and both
$\cal D$ and $\cal Z$ are dense in $\cal S$. One proves
\cite{Gelfand-S} that the Fourier transformation which defines the
unitary operator
\begin{eqnarray} F\, :\, L^2(\bbR)\, \longrightarrow\, L^2(\bbR)\
, \end{eqnarray} establishes an isomorphism between $\cal D$ and
$\cal Z$.

\section{Quantization of damped systems}

Let us  quantize a classical damped system described by
(\ref{damp}). Clearly this system is not Hamiltonian. However, it
is well known (cf.  \cite{Pontr}) that  any dynamical system may
be rewritten in a Hamiltonian form. Consider a dynamical system on
$n$-dimensional configuration space $Q$:
\begin{equation}\label{dotx-X}
  \dot{x} = X(x)\ ,
\end{equation}
where $X$ is a vector field on $Q$. Now, define the following
Hamiltonian on the cotangent bundle ${\cal P}=T^*Q$:
\begin{equation}\label{}
  H(\alpha_x) := \alpha_x(X(x))\ ,
\end{equation}
where $\alpha_x \in T_x^*Q$. Using canonical coordinates
$(x^1,\ldots,x^n,p_1,\ldots,p_n)$ one obtains:
\begin{equation}\label{}
  H(x,p) = \sum_{k=1}^n p_kX^k(x)\ ,
\end{equation}
where $X^k$ are components of $X$ in the coordinate basis
$\partial/\partial x^k$. The corresponding Hamilton equations take
the following form:
\begin{eqnarray}\label{HAM-1}
  \dot{x}^k &=& \{ x^k,H\} = X^k(x) \ , \\
  \dot{p}_k &=& \{ p_k,H\} = - \sum_{l=1}^n p_l \frac{\partial
  X^l(x)}{\partial x^k} \ ,
\end{eqnarray}
for $k=1,\ldots,n$. In the above formulae $\{\ , \ \}$ denotes the
canonical Poisson bracket on $T^*Q$:
\begin{equation}\label{}
  \{F,G\} = \sum_{k=1}^n \left( \frac{\partial F}{\partial x^k}
   \frac{\partial G}{\partial p_k} - \frac{\partial G}{\partial x^k}
   \frac{\partial F}{\partial p_k}  \right) \ .
\end{equation}
Clearly, the formulae (\ref{HAM-1}) reproduce our initial
dynamical system (\ref{dotx-X}) on $Q$.

Let us apply the above procedure to the damped system
(\ref{damp}). One obtains for the Hamiltonian
\begin{equation}\label{}
  H(x,p) = - \gamma xp \ ,
\end{equation}
and hence the corresponding Hamilton equations
\begin{eqnarray}\label{}
\dot{x} = - \gamma x \ , \hspace{1cm} \dot{p} = \gamma p \ ,
\end{eqnarray}
give rise to the following Hamiltonian flow on $ \mathbb{R}^2$:
\begin{equation}\label{}
  (x,p) \ \longrightarrow \ (e^{-\gamma t}x,e^{t\gamma}p)\ .
\end{equation}
Now, the quantization is straightforward: one has for the Hilbert
space ${\cal H} = L^2(\bbR)$, and for the Hamiltonian
\begin{eqnarray} \label{Ham} \widehat{H} = - \frac{\gamma}{2}
(\widehat{x}\widehat{p} + \widehat{p}\widehat{x}) \ .
\end{eqnarray} It is evident that (\ref{Ham}) defines a symmetric
operator on $L^2(\bbR)$. In section~\ref{PROPERTIES} we show that
$\widehat{H}$ is self-adjoint and hence it gives rise to a well
defined quantum mechanical problem.\footnote{Actually, this
Hamiltonian is well known in quantum optics in connection with the
squeezed states of light \cite{Walls}. Introducing $\widehat{a}$
and $\widehat{a}^*$: \begin{eqnarray} \widehat{x} =
\frac{\widehat{a} + \widehat{a}^*}{\sqrt{2}}\ , \ \ \ \ \ \ \
\widehat{p} = \frac{\widehat{a} - \widehat{a}^*}{\sqrt{2}i}\
,\nonumber \end{eqnarray} the Hamiltonian (\ref{Ham}) may be
rewritten as follows: \begin{eqnarray} \widehat{H} =
\frac{\gamma}{2i} \left( \widehat{a}^{*2} - \widehat{a}^2\right) \
,\nonumber \end{eqnarray} which is exactly a generator of
squeezing.     }

Let us observe that performing the canonical transformation
\begin{eqnarray} x = \frac{1}{\sqrt{2\gamma}}\, (  \gamma  X - P
)\ ,       \hspace{1cm} p = \frac{1}{\sqrt{2\gamma}}\, (  \gamma X
+ P )\ ,
\end{eqnarray} the classical Hamiltonian (\ref{Ham}) takes the
following form: \begin{eqnarray} \widehat{H} = \frac{1}{2} (
\widehat{P}^2 - \gamma ^2 \widehat{X}^2) \ , \end{eqnarray} that
is, it corresponds to the so called reversed harmonic oscillator.
This system was  analyzed in \cite{LL} and recently in
\cite{Ann1,Ann2,Castagnino} (see also \cite{Koopman,JPA}).

\section{Properties of the Hamiltonian}
\label{PROPERTIES}

Let us investigate the basic properties of the Hamiltonian defined
in (\ref{Ham}).

\begin{proposition}
The operator $\widehat{H} = - \frac{\gamma}{2}
(\widehat{x}\widehat{p} + \widehat{p}\widehat{x} )$ is
self-adjoint in $ L^2(\mathbb{R})$.
\end{proposition}

\noindent {\em Proof.} To prove that $\widehat{H}$ is self-adjoint
we show that $e^{-i\widehat{H}}$ is unitary in $L^2(\mathbb{R})$.
One has \begin{eqnarray} \widehat{H} = - \frac{\gamma}{2}
(\widehat{x}\widehat{p} + \widehat{p}\widehat{x}) = i\gamma \left(
x\frac{d}{dx} + \frac 12 \right) \ . \end{eqnarray} Let us define
\begin{eqnarray} U = e^{-i\widehat{H}} = e^{\gamma/2}e^{\gamma x
\partial_x}\ . \end{eqnarray} Clearly, \begin{eqnarray} \label{U-gamma} U\psi(x) =
e^{\gamma/2}\psi(e^{\gamma}x) \ , \end{eqnarray} for any $\psi \in
L^2(\mathbb{R})$. The operator $U$ defines an isometry:
\begin{eqnarray}
\la U\psi|U\phi\r &=& \int_{-\infty}^\infty
\overline{U\psi(x)}U\phi(x)\, dx = \int_{-\infty}^\infty e^\gamma
\overline{\psi(e^\gamma x)} \phi(e^\gamma x) \, dx =
\int_{-\infty}^\infty \overline{\psi(y)}\phi(y)\, dy \nonumber \\
&=&  \la \psi|\phi\r \ .
\end{eqnarray}
Moreover, due to (\ref{U-gamma}), $U$ is onto, and hence it is
unitary in   $L^2(\mathbb{R})$. Therefore,  Stone's theorem
implies that $\widehat{H}$ is self-adjoint (see e.g.
\cite{Yosida}). \hfill $\Box$

\noindent Obviously, $\widehat{H}$ is parity invariant:
\begin{eqnarray}
 {\bf P}\widehat{H} {\bf P}^{-1} = \widehat{H}\ ,
\end{eqnarray} where the parity operator $P$ is defined by: \begin{eqnarray}
{\bf P}\widehat{x} {\bf P}^{-1} = -\widehat{x}\ , \hspace{1cm}
{\bf P}\widehat{p} {\bf P}^{-1} = -\widehat{p}\ . \end{eqnarray}
Now, let us turn to the time reversal operator {\bf T}. The theory
invariant under the time reversal has the following property: if
$\psi(t)$ is a solution of the Schr\"odinger equation  given by
\begin{eqnarray} \psi(t) = U(t)\psi \ , \end{eqnarray} with $U(t) =
e^{-i\widehat{H}t}$, then ${\bf T}\psi$ evolves into
\begin{eqnarray} ({\bf T}\psi)(-t) = U(t)({\bf T}\psi)\ , \end{eqnarray} or,
equivalently \begin{eqnarray} \label{T-U} {\bf T}(U(t)\psi) =
U(-t)({\bf T}\psi)\ , \end{eqnarray} for any $\psi \in {\cal H}$.
Now, following Wigner \cite{Wigner}, {\bf T} is either unitary or
anti-unitary. If {\bf T} is unitary, then (\ref{T-U}) implies
\begin{eqnarray} {\bf T}\widehat{H} + \widehat{H}{\bf T} = 0\ .
\end{eqnarray} It means that if \begin{eqnarray} \widehat{H}\psi^E=E\psi^E\
, \end{eqnarray} then \begin{eqnarray} \widehat{H}\, {\bf
T}\psi^E= - E\, {\bf T}\psi^E\ , \end{eqnarray} that is, any
eigenvector $\psi^E$ with the energy $E$ is accompanied by ${\bf
T}\psi^E$ with energy $-E$. Usually, this case is excluded since
one expects that the Hamiltonian is bounded from below. If this is
the case, then $\bf T$ is anti-unitary and (\ref{T-U}) implies:
\begin{eqnarray} {\bf T}\widehat{H} - \widehat{H}{\bf T} = 0\ .
\end{eqnarray} However, the Hamiltonian defined in (\ref{Ham}) is not
bounded from below, and, as we show in section~\ref{Spectrum} its
spectrum $\sigma(\widehat{H})=(-\infty,\infty)$. Therefore, we
take {\bf T} to be unitary in $L^2(\bbR)$.

\begin{proposition} The time reversal operator {\bf T} is realized by the Fourier transformation:
\begin{eqnarray} {\bf T}\psi := F[\psi]\ , \end{eqnarray} i.e. \begin{eqnarray} F^{-1}\widehat{H}F
\psi = - \widehat{H}\psi\ , \end{eqnarray} for all $\psi \in
L^2(\bbR)$. Moreover, \begin{eqnarray} {\bf T}^2\psi(x) = {\bf
P}\phi(x) = \psi(-x) \ . \end{eqnarray}
\end{proposition}
Denoting by $\bf C$ the complex conjugation ${\bf C}\psi =
\overline{\psi}$, one immediately finds

\begin{proposition} The Hamiltonian (\ref{Ham}) is {\bf CT} and {\bf PCT}
invariant, i.e. \begin{eqnarray} [\widehat{H},{\bf CT}] =
[\widehat{H},{\bf PCT}]=0\ . \end{eqnarray}
\end{proposition}
Therefore, if \begin{eqnarray} \widehat{H}\psi^E = E\psi^E\ ,
\end{eqnarray} then \begin{eqnarray} \label{CT}
\widehat{H}F[\overline{\psi^E}] = EF[\overline{\psi^E}]\ .
\end{eqnarray} Clearly, {\bf CT} invariance  does not produce any
conserved quantity since {\bf CT} is anti-unitary.

\section{Complex eigenvalues}
\label{Complex}

Interestingly, $\widehat{H}$ being  self-adjoint admits
generalized eigenvectors with complex eigenvalues
\cite{Castagnino,Kossak,Koopman,JPA}.
 Let $f^\pm_0$ be distributions satisfying
\begin{eqnarray} \widehat{x}\, f^-_0 = 0 \ , \hspace{1cm} \widehat{p}\, f^+_0=
0 \ . \end{eqnarray} Clearly, \begin{eqnarray} f^-_0(x) =
\delta(x)\ , \hspace{1cm} f^+_0(x)=1\ . \end{eqnarray} Its easy to
see that
\begin{eqnarray} \widehat{H}\, f^\pm_0 = \pm i \frac \gamma 2 \,
f^\pm_0\ . \end{eqnarray} Let us define two families:
\begin{eqnarray} f^-_n := \frac{(-i)^n}{\sqrt{n!}}\,
\widehat{p}^n\, f^-_0\ , \hspace{1cm} f^+_n :=
\frac{1}{\sqrt{n!}}\, \widehat{x}^n\, f^+_0\ . \end{eqnarray} One
finds
\begin{eqnarray}  \label{fn} f^-_n(x) = \frac{(-1)^n}{\sqrt{n!}}\,
\delta^{(n)}(x)\ , \hspace{1cm} f^+_n(x) = \frac{x^n}{\sqrt{n!}}\
. \end{eqnarray} Moreover,
\begin{eqnarray} \widehat{H}\, f^\pm_n = \pm E_n \, f^\pm_n\ ,
\end{eqnarray} where \begin{eqnarray} \label{En} E_n := i\gamma \left( n +
\frac 12 \right) \ . \end{eqnarray} Clearly, both $f^-_n$ and
$f^-_n$ are tempered distributions, i.e. $f^\pm_n \in {\cal S}'$.
Evidently, they are related by the Fourier transformation:
\begin{eqnarray}
F[f^+_n] = \sqrt{2\pi} i^n f^-_n\ , \hspace{1cm} F[f^-_n] =
\frac{i^n}{\sqrt{2\pi}} f^+_n\ .
\end{eqnarray}
 Let us observe, that these two families of generalized eigenvectors have two remarkable properties:
\begin{eqnarray}   \label{P1} \int_{-\infty}^\infty f^+_n(x)\, f^-_m(x)\, dx =
\delta_{nm}\ , \end{eqnarray} and \begin{eqnarray}   \label{P2}
\sum_{n=0}^\infty\, f^+_n(x)\, f^-_n(x') = \delta(x-x')\ .
\end{eqnarray} These formulae remind one of the basic basic
properties of proper (Hilbert space) eigenvectors: if
$\widehat{A}$ is a self-adjoint operator in $\cal H$ and
\begin{eqnarray} \widehat{A}\psi_k = \lambda_k \psi_k\ , \end{eqnarray}
where $\psi_k$ are normalized vectors in $\cal H$, then
\begin{eqnarray} \int \overline{\psi_n}(x)\psi_m(x)\, dx =
\delta_{nm}\ , \end{eqnarray} and
\begin{eqnarray} \sum_n \overline{\psi_n}(x)\psi_n(x')\, dx =
\delta(x-x')\ . \end{eqnarray} Obviously, there is no complex
conjugation in (\ref{P1}) and (\ref{P2}) since $f^\pm_n$ are real
functions.

Now, for any $\phi \in {\cal Z}$ one has
\begin{eqnarray}
\phi(x) = \sum_n \frac{\phi^{(n)}(0)}{n!} (-1)^nx^n = \sum_n
f^+_n(x) \la f^-_n|\phi\r \ .
\end{eqnarray}
On the other hand, for any $\phi \in {\cal D}$, its Fourier
transform $F[\phi] \in {\cal Z}$, and
\begin{eqnarray}
\phi(x) &=& \frac{1}{\sqrt{2\pi}} \int e^{ikx} F[\phi](k) dk =
\frac{1}{\sqrt{2\pi}} \int e^{ikx} \sum_n \frac{ F[\phi]^{(n)}(0)}{n!} (-1)^n k^n \, dk \nonumber \\
&=& \sum_n F[f^+_n](x) \la f^-_n|F[\phi]\r = \sum_n F[f^+_n](x)
\la F[f^-_n]|\phi\r \nonumber \\ &=& \sum_n f^-_n(x) \la
f^+_n|\phi\r \ .
\end{eqnarray}
Hence, we have two spectral decompositions:

\begin{eqnarray}  \label{Z} |\phi\r =  \sum_n |f^+_n\r \la f^-_n|\phi\r
\hspace{1cm} \mbox{in}\ \ \ {\cal Z} \ , \end{eqnarray} and
\begin{eqnarray} \label{D} |\psi\r =  \sum_n |f^-_n\r \la
f^+_n|\psi\r \hspace{1cm} \mbox{in}\ \ \ {\cal D}\ .
\end{eqnarray} In section \ref{ANA} we derive (\ref{Z}) and
(\ref{D}) from the spectrum of $\widehat{H}$. So let us look for
$\sigma(\widehat{H})$.

\section{Spectrum}
\label{Spectrum}

The Hamiltonian (\ref{Ham}) has a continuous spectrum
$\sigma(\widehat{H}) = (-\infty,\infty)$. Since, the Hamiltonian
(\ref{Ham}) is parity invariant
 each generalized eigenvalue $E \in  \bbR$ is doubly degenerated:
\begin{eqnarray} \widehat{H} \psi^E_\pm = E\psi^E_\pm\ . \end{eqnarray} The above
equation may be rewritten as the following differential equation
for $\psi^E_\pm$: \begin{eqnarray}  \label{eq}
x\frac{d}{dx}\psi^E_\pm(x) = - \left( i \frac{E}{\gamma} + \frac
12 \right) \psi^E_\pm \ . \end{eqnarray} To solve (\ref{eq}) let
us introduce the following distributions \cite{Gelfand-S} (see
also \cite{Kanwal}): \begin{eqnarray} x^\lambda_+ := \left\{
\begin{array}{ll} x^\lambda & \ \ \ x\geq 0 \\ 0 &\ \ \ x<0
\end{array} \right. \ , \hspace{1cm} x^\lambda_- := \left\{
\begin{array}{cl} 0 & \ \ \  x\geq0 \\ |x|^\lambda  &\ \ \ x<0
\end{array} \right. \ , \end{eqnarray} with $\lambda \in \bbC$ (basic
properties of $x^\lambda_\pm$ are collected in the Appendix). It
is,  therefore, clear  that the generalized eigenvectors
$\psi^E_\pm$ may be written as follows: \begin{eqnarray}
\label{psi-E} \psi^E_\pm(x) := \frac{1}{\sqrt{2\pi \gamma}} \,
x^{-(iE/\gamma + 1/2)}_\pm\ . \end{eqnarray} It turns out that
$\psi^E_\pm$ are well defined tempered distributions for all $E
\in \bbR$. Actually, instead of $\psi^E_\pm$ one may work with
eigenvectors of the parity operator $\bf P$:
\begin{eqnarray}
\psi^E_{\rm even} &=& \frac{1}{\sqrt{2}} \left( \psi^E_+ + \psi^E_- \right) \ ,\\
\psi^E_{\rm odd} &=& \frac{1}{\sqrt{2}} \left( \psi^E_+ - \psi^E_-
\right) \ .
\end{eqnarray}
Obviously
\begin{eqnarray} {\bf P}\, \psi^E_{\rm even} = \psi^E_{\rm even}\ , \hspace{1cm}
 {\bf P}\, \psi^E_{\rm odd} = - \psi^E_{\rm odd}\ .
\end{eqnarray}
These distributions of definite parity are given by:
\begin{eqnarray}
 \psi^E_{\rm even} = \frac{1}{2\sqrt{\pi \gamma}} \, |x|^{-(iE/\gamma + 1/2)}\ , \hspace{1cm}
\psi^E_{\rm odd} = \frac{1}{2\sqrt{\pi \gamma}} \, {\rm
sign}(x)|x|^{-(iE/\gamma + 1/2)}\ ,
\end{eqnarray}
(see \cite{Gelfand-S} and \cite{Kanwal} for the properties of
$|x|^\lambda$ and ${\rm sign}(x)|x|^\lambda$).

With the normalization used in (\ref{psi-E})  one proves
\cite{Bollini} orthonormality:
\begin{eqnarray}
\int \overline{\psi^{E_1}_\pm(x)}\psi^{E_2}_\pm(x)\, dx =
\delta(E_1-E_2) \ ,
\end{eqnarray}
and completeness:
\begin{eqnarray}
\int \overline{\psi^{E}_\pm(x)}\psi^{E}_\pm(x')\, dE =
\delta(x-x') \ .
\end{eqnarray}
 Therefore, due to the Gelfand-Maurin spectral theorem one has
\begin{eqnarray}   \label{GM-1} \phi(x) = \sum_\pm \int dE\, \psi^E_\pm(x) \la
\psi^E_\pm|\phi\r \ , \end{eqnarray} for any $\phi \in {\cal S}$,
and the corresponding  spectral resolution of the Hamiltonian has
the following form: \begin{eqnarray}   \label{GMH-1} \widehat{H} =
\sum_\pm \int dE\, E | \psi^E_\pm \r \la \psi^E_\pm| \ .
\end{eqnarray} There is another family of energy eigenvectors
directly related to $\psi^E_\pm$. Due to (\ref{CT}) one has:
\begin{eqnarray} \widehat{H}\, F[{\psi}^{-E}_\pm] = E
F[{\psi}^{-E}_\pm]\ . \end{eqnarray} The Fourier transform of
$\psi^E_\pm$ is given by
  (cf. \cite{Gelfand-S} and the Appendix):
\begin{eqnarray}
F[\psi^{-E}_\pm](k) = \pm \frac{i}{2\pi \sqrt{\gamma}} \exp\left[
\pm \frac{i\pi}{2} \left( i \frac E \gamma - \frac 12 \right)
\right] \Gamma\left( i \frac E \gamma + \frac 12 \right)
 (k \pm i0)^{-(iE/\gamma + 1/2)}\ .
\end{eqnarray} One shows \cite{Gelfand-S}  that  $F[\psi^E_\pm]$ are  well
defined tempered distributions for any $E \in \bbR$. Moreover,
\begin{eqnarray} \int
\overline{F[{\psi}^{E_1}_\pm]}(x)\,F[{\psi}^{E_2}_\pm](x)\, dx =
\delta(E_1-E_2) \ , \end{eqnarray} and \begin{eqnarray} \int
\overline{F[{\psi}^{E}_\pm]}(x)\,F[{\psi}^{E}_\pm](x')\, dE =
\delta(x-x') \ . \end{eqnarray} Hence,  following the
Gelfand-Maurin theorem, we have further spectral decompositions:
for any $\psi \in {\cal S}$ \begin{eqnarray} \label{GM-2} \psi(x)
= \sum_\pm \int dE\, F[{\psi}^{-E}_\pm](x) \la
F[{\psi}^{-E}_\pm]|\psi\r \ , \end{eqnarray}
 and for the Hamiltonian itself:
\begin{eqnarray}    \label{GMH-2} \widehat{H} =  \sum_\pm \int dE\, E |
F[{\psi}^{-E}_\pm] \r \la F[{\psi}^{-E}_\pm]| \ . \end{eqnarray}

\section{Analyticity of energy eigenvectors}
\label{ANA}

Let us continue the energy eigenvectors $\psi^E_\pm$ and
$F[\psi^{-E}_\pm]$ into the energy complex plane $E \in \bbC$. It
turns out \cite{Gelfand-S} (see also the Appendix) that
$\psi^E_\pm$ has  simple poles at $E=-E_n$, whereas
$F[\psi^{-E}_\pm]$ has simple poles at $E=+E_n$, with $E_n$
defined in (\ref{En}). Therefore, the poles of energy eigenvectors
considered as functions of the complex energy correspond exactly
to the complex eigenvalues of $\widehat{H}$ which we found in
Section~\ref{Complex}. One easily computes the corresponding
residues: \begin{eqnarray}  \label{Res-1}
\mbox{Res}(\psi^E_\pm(x);-E_n) = i\, (\mp 1)^n \sqrt{
\frac{\gamma}{2\pi}}\frac{\delta^{(n)}(x)}{n!}  \ , \end{eqnarray}
and \begin{eqnarray}  \label{Res-2}
\mbox{Res}(F[\psi^{-E}_\pm(x)];+E_n) = \pm
\frac{\sqrt{\gamma}}{2\pi} (\mp i)^{n+1} \frac{(-1)^n}{n!} x^n\ .
\end{eqnarray} Hence, residues of $\psi^E_\pm$ and $F[\psi^{-E}_\pm]$
correspond, up to numerical factors, to the eigenvectors $f^\pm_n$
(\ref{fn}): \begin{eqnarray} \mbox{Res}(\psi^E_\pm(x);-E_n) \ \sim
\ f^-_n \ , \end{eqnarray} and
\begin{eqnarray} \mbox{Res}(F[\psi^{-E}_\pm(x)];+E_n) \ \sim \
f^+_n \ . \end{eqnarray} Any function $\phi \in  {\cal S} \subset
L^2(\bbR)$ gives rise to the following functions of energy:
\[  \bbR \ni E\ \longrightarrow \ \la \psi^E_\pm|\phi\r \in \bbC\ , \]
and
\[  \bbR \ni E\ \longrightarrow \ \la F[\psi^{-E}_\pm]|\phi\r \in \bbC\ . \]
Let us introduce two important classes of functions  \cite{Duren}:
a smooth function $f=f(E)$ is in the  Hardy class from above
${\cal H}^2_+$ (from below ${\cal H}^2_-$) if $f(E)$ is a boundary
value of an analytic function  in the upper, i.e. $\mbox{Im}\,
E\geq 0$ (lower, i.e. $\mbox{Im}\, E\leq 0$) half complex
$E$-plane vanishing faster than any power of $E$ at the upper
(lower) semi-circle $|E| \rightarrow \infty$. Now, define
\begin{eqnarray} \Phi_- := \Big\{ \phi \in {\cal S}\, \Big| \, \la
\psi^E_\pm | \phi \r \in {\cal H}^2_-\, \Big\} \ , \end{eqnarray}
and
\begin{eqnarray} \Phi_+ := \Big\{ \phi \in {\cal S}\, \Big| \, \la
F[\psi^{-E}_\pm] | \phi \r \in {\cal H}^2_+\, \Big\} \ .
\end{eqnarray}

\begin{proposition} $ \Phi_+ \cap \Phi_- = \{\emptyset\}$.
\end{proposition}

\noindent {\em Proof.} Clearly, if  $\phi \in \Phi_-$, then $\la
\psi^E_\pm|\phi\r$ is a smooth function of $E\in \bbR$. Suppose,
that $\phi \in \Phi_+$, that is \begin{eqnarray} \la
F[\psi^{-E}_\pm]|\phi \r = \la \psi^{-E}_\pm|F[\phi]\r \in {\cal
H}^2_+ \ . \end{eqnarray} Now, due to the Paley-Wiener theorem
\cite{Yosida} the inverse Fourier transform of $F[\phi]$
\begin{eqnarray} {F}^{-1}[F[\phi]](E) = \frac{1}{\sqrt{2\pi}}
\int_{-\infty}^\infty F[\phi](t)\, e^{-itE}\, dt\ , \end{eqnarray}
vanishes for $E >0$. Therefore, $\phi(E)=0$ for $E>0$, and hence
$f(E)$ cannot be a smooth function of $E$. \hfill $\Box$

\noindent Our main result consists in the following
\begin{theorem} For any $\phi^\pm \in \Phi_\pm$ one has
\begin{eqnarray}  \label{phi-} \phi^-(x) = \sum_n f^-_n(x) \la f^+_n|\phi^-\r
\ , \end{eqnarray} and \begin{eqnarray}   \label{phi+} \phi^+(x) =
\sum_n f^+_n(x) \la f^-_n|\phi^+\r \ . \end{eqnarray}
\end{theorem}

\noindent {\em Proof.} Due to the spectral formula (\ref{GM-1})
one has, for $\phi^- \in \Phi_- \subset {\cal S}$:
\begin{eqnarray} \phi^-(x) = \sum_\pm \int_{-\infty}^\infty dE\,
\psi^E_\pm(x) \la \psi^E_\pm | \phi^-\r \ . \end{eqnarray} Now,
since $ \la \psi^E_\pm | \phi^-\r \in {\cal H}^2_-$, we may close
the integration contour along the lower semi-circle
$|E|\rightarrow \infty$. Hence, due to the residue theorem one
obtains
\begin{eqnarray} \label{phi-R} \phi^-(x) = -2\pi i \sum_\pm \sum_n
\mbox{Res}(\psi^E_\pm(x);-E_n)\, \la \psi^{E}_\pm |
\phi^-\r\Big|_{E=-E_n} \ . \end{eqnarray} Using the definition of
$\psi^E_\pm$ \begin{eqnarray} \la \psi^E_\pm | \phi^-\r =
\frac{1}{\sqrt{2\pi\gamma}} \int \overline{ x^{-(iE/\gamma +
1/2)}_\pm} \, \phi^-(x) dx = \frac{1}{\sqrt{2\pi\gamma}} \int
x^{-(-iE/\gamma + 1/2)}_\pm \, \phi^-(x)\, dx\ , \end{eqnarray}
one finds
\begin{eqnarray}
 \la \psi^{E}_\pm | \phi^-\r\Big|_{E=-E_n} = \frac{1}{\sqrt{2\pi\gamma}} \int x^n_\pm \phi^-(x)\, dx\ .
\end{eqnarray} Therefore, inserting into (\ref{phi-R}) the value of the
residue given in  (\ref{Res-1}) one gets finally
\begin{eqnarray}
\phi^-(x) &=& \sum_n \frac{\delta^{(n)}(x)}{n!} \, \int \left[
(-1)^nx^n_+ + x^n_-\right] \phi^-(x)\, dx = \sum_n
(-1)^n\frac{\delta^{(n)}(x)}{n!} \, \int x^n\phi^-(x)\, dx
\nonumber \\ &=&
 \sum_n f^-_n(x) \la f^+_n|\phi^-\r \ .
\end{eqnarray}
To prove (\ref{phi+}) let us use another spectral formula
(\ref{GM-2}): for any $\phi^+ \in \Phi_+ \subset {\cal S}$
\begin{eqnarray}
\phi^+(x) = \sum_\pm \int_{-\infty}^\infty dE\,
F[\psi^{-E}_\pm](x) \la F[\psi^{-E}_\pm] | \phi^-\r \ .
\end{eqnarray}
Now, since $ \la F[\psi^{-E}_\pm] | \phi^-\r \in {\cal H}^2_+$, we
may close the integration contour along the upper semi-circle
$|E|\rightarrow \infty$. Hence the residue theorem implies
\begin{eqnarray}  \label{phi+R}
\phi^+(x) = +2\pi i \sum_\pm \sum_n
\mbox{Res}(F[\psi^{-E}_\pm(x)];+E_n)\, \la F[\psi^{-E}_\pm] |
\phi^+\r\Big|_{E=+E_n} \ .
\end{eqnarray}
Now, using once more the formula for $\psi^E_\pm$ one finds
\begin{eqnarray}
\la F[\psi^{-E}_\pm] | \phi^+\r\Big|_{E=+E_n} =
\frac{1}{\sqrt{2\pi \gamma}} \la F[x^n_\pm]|\phi^+\r \ .
\end{eqnarray}
Hence, inserting the values of residues (\ref{Res-2}) into
(\ref{phi+R}) and using the formula for $F[x^n_\pm]$ (see
(\ref{F-x-n})) one has
\begin{eqnarray}
\phi^+(x) &=& \frac{i}{\sqrt{2\pi}} \sum_n (-1)^n \frac{x^n}{n!}
\, \Big[ (-i)^{n+1} \la F[x^n_+]|\phi^+\r - i^{n+1} \la
F[x^n_-]|\phi^+\r \Big] \nonumber \\ &=& \frac{i}{\sqrt{2\pi}}
\sum_n (-1)^n \frac{x^n}{n!}\, \int \left[ (-i)^{n+1}  \overline{
F[x^n_+](k)} - i^{n+1} \overline{ F[x^n_-](k)}\right] \phi^+(k)\,
dk
  \nonumber \\ &=&
\frac{i}{{2}} \sum_n (-1)^n \frac{x^n}{n!} \, \Big[ (-i)^{n+1}i^n
- i^{n+1}(-i)^n \Big] \int \delta^{(n)}(k)\phi^+(k)\, dk\nonumber
\\ & =&  \sum_n f^+_n(x) \la f^-_n|\phi^+\r \ ,
\end{eqnarray}
which ends the proof.  \hfill $\Box$

\noindent This way we have recovered (\ref{Z}) and (\ref{D}). It
is not surprising, due to the following

\begin{proposition} $\Phi_- = {\cal Z}$ and $\Phi_+ = {\cal D}$.
\end{proposition}

\begin{corollary} We have two spectral decomposition of $\widehat{H}$:
\begin{eqnarray}   \label{H1} \widehat{H} = \sum_n \overline{E}_n |f^-_n\r\la
f^+_n|   \hspace{1cm} \rm{on}\ \ \ \Phi_-\ , \end{eqnarray} and
\begin{eqnarray} \label{H2} \widehat{H} = \sum_n {E}_n |f^+_n\r\la
f^-_n| \hspace{1cm} \rm{on}\ \ \ \Phi_+\ . \end{eqnarray}
\end{corollary}

\section{Resonances and the quantum damping}

Finally, let us turn to the evolution generated by the Hamiltonian
(\ref{Ham}). Obviously, it generates a 1-parameter unitary group
\begin{eqnarray} U(t) = e^{-i\widehat{H}t}\ , \end{eqnarray} on $L^2(\bbR)$. It follows
from (\ref{U-gamma}) that \begin{eqnarray} \psi_t(x) = U(t)\psi(x)
= e^{\gamma t/2}\psi(e^{\gamma t}x)\ . \end{eqnarray} The above
formula is well defined for any $t\in \bbR$ and clearly, as we
already showed, the theory is time-reversal invariant: if
$\psi(t)$ solves the Schr\"odinger equation so does ${\bf
T}\psi(t):=\psi(-t)$. Therefore, working in $L^2(\bbR)$ we do not
see any damping at all. Now, let us construct two natural Gelfand
triplets:
\begin{eqnarray}
\Phi_\pm \subset L^2(\bbR) \subset \Phi_\pm '\ .
\end{eqnarray}
If $\phi^- \in \Phi_-$, then
\begin{equation}\label{}
  \la \psi^E_\pm | U(t)\phi^-\r =  \la U^*(t)\,\psi^E_\pm  |\phi^-\r
  = e^{-iEt}\,  \la \psi^E_\pm |\phi^-\r \ .
\end{equation}
Hence $\phi^-(t)\in \Phi_-$ only for $t\geq 0$. Similarly, if
$\phi^+ \in \Phi_+$, then $\phi^+(t)\in \Phi_+$ only for $t\leq
0$. Therefore, the restriction of the unitary group $U(t)$ on
$L^2( \mathbb{R})$ to $\Phi_\pm$ no longer defines a group. It
gives rise to two semigroups:
\begin{equation}
 U_-(t) \ :\ \Phi_-
\ \longrightarrow\ \Phi_-\ ,\ \ \ \ \ \ {\rm for}\ \ \ t\geq 0\ ,
\end{equation}
and
\begin{equation}
  U_+(t) \ :\ \Phi_+ \ \longrightarrow\ \Phi_+\ ,\ \ \ \ \ \
{\rm for}\ \ \ t\leq 0\ .
\end{equation}
 Due to (\ref{H1}) and (\ref{H2}) one has:
\begin{eqnarray} \label{phi:-}
\phi^-(t) = U(t)\phi^- = \sum_n e^{-\gamma(n+1/2)t} |f^-_n\r\la
f^+_n|\phi^-\r \ ,
\end{eqnarray}
for $t\geq 0$, and
\begin{eqnarray} \label{phi:+}
\phi^+(t) = U(t)\phi^+ = \sum_n
e^{\gamma(n+1/2)t} |f^+_n\r\la f^-_n|\phi^+\r \ ,
\end{eqnarray}
for $t\leq 0$.  We stress that $\phi^-_t$ ($\phi^+_t$) does belong
to $L^2(\bbR)$ also for $t<0$ ($t>0$). However, $\phi^-_t \in
\Phi_-$ ($\phi^+_t \in \Phi_+$) only for $t\geq 0$ ($t\leq 0$).
This way the irreversibility enters on a purely Hamiltonian level
by restricting dynamics to the dense subspace $\Phi_\pm$ of
$L^2(\bbR)$.

Clearly, formulae (\ref{phi:-}) and (\ref{phi:+}) are quantum
analogues of the classical damping laws:
\begin{eqnarray}   x(t) = e^{-\gamma t} x\  , \hspace{1cm} t\geq 0
\ ,
\end{eqnarray}
and
\begin{eqnarray}
 p(t) = e^{+\gamma t}p\ , \hspace{1cm} t\leq 0
\ .
\end{eqnarray}
Finally, let us recall that the time reversal operator {\bf T}
establishes an isomorphism between $\Phi_-$ and $\Phi_+$.
Therefore, each solution
\begin{eqnarray}
\phi^-_t = U_-(t)\phi^-\ ,
\end{eqnarray}
with $\phi^- \in \Phi_-$ is mapped into
\begin{eqnarray}
{\bf T}(\phi^-_t) = U_-(-t){\bf T}(\phi^-) = U_+(t){\bf
T}(\phi^-)\ , \ \ \ \ t\leq 0 \ .
\end{eqnarray}
Conversely, any solution
\begin{eqnarray}
\phi^+_t = U_+(t)\phi^+\ ,
\end{eqnarray}
with $\phi^+ \in \Phi_+$ is mapped into
\begin{eqnarray}
{\bf T}(\phi^+_t) = U_+(-t){\bf T}(\phi^+) =U_-(t){\bf T}(\phi^+)\
, \ \ \ \ t\geq 0 \ .
\end{eqnarray}
Summarizing, quantum dynamics is irreversible on $\Phi_-$ and
$\Phi_+$. This irreversibility is caused by quantum damping, or,
equivalently, by the presence of resonant states $f^\pm_n$
(\ref{fn}).  It should be stressed that it is not an energy that
is dissipated. Clearly, the Hamiltonian (\ref{Ham}) can not be
interpreted as a system energy --- it was used to define a
Hamiltonian dynamics of the enlarged system on $L^2( \mathbb{R})$.
The quantum damped system is not defined on the entire $L^2(
\mathbb{R})$ but rather on a dense subset $\Phi_- \subset L^2(
\mathbb{R})$. As we saw it imposes the restriction upon the time
evolution such that it is defined only for positive $t$. A quantum
damping  may be seen as follows: let $\phi^-_0 \in \Phi_-$ be an
initial state then the probability density for a particle position
evolves in time as follows:
\begin{equation}\label{}
  p_t(x) = |\phi^-_t(x)|^2 = e^{\gamma t}|\phi^-_0(e^{\gamma
  t}x)|^2  = e^{\gamma t} p_0(e^{\gamma t}x) \ ,
\end{equation}
and hence in the limit $t \longrightarrow +\infty$, one finds
$p_t(x) \longrightarrow \delta(x)$. Indeed, for any $\epsilon >0$
\begin{equation}\label{}
  \int_{-\epsilon}^\epsilon p_t(x)dx =
   \int_{-\epsilon e^{\gamma t}}^{\epsilon e^{\gamma t}} p_0(x)dx \
   \longrightarrow\ 1 \ ,
\end{equation}
for $t \longrightarrow +\infty$. Clearly, it corresponds to the
classical behavior $x(t) = e^{-\gamma t}x_0\longrightarrow 0$.

In a forthcoming paper we are going to show that also more
complicated damped systems, e.g. the damped harmonic oscillator,
give rise to irreversible dynamics.

\appendix

\section{Appendix}

The regular tempered distribution $x^\lambda_+$ (with $\lambda \in
\mathbb{C}$) given by
\begin{eqnarray} \la \phi|x^\lambda_+\r :=
\int_0^{\infty} x^\lambda \phi(x)dx \ , \end{eqnarray}
 for any
$\phi \in {\cal S}$, is well defined for $\mbox{Re}\,\lambda >
-1$. However, it may be easily extended to the region
$\mbox{Re}\lambda > -2$ due to the following regularization
formula:
\begin{eqnarray} \int_0^{\infty} x^\lambda \phi(x)dx =
\int_0^{1} x^\lambda [\phi(x)-\phi(0)]dx + \int_1^{\infty}
x^\lambda \phi(x)dx + \frac{\phi(0)}{\lambda +1} \ ,
\end{eqnarray}
which holds for $\lambda \neq -1$. In the same way one may extend
the distribution $x^\lambda_+$ to   the region $\mbox{Re}\,
\lambda > -n-1$ using the formula
\begin{eqnarray}    \label{pole-1}
\int_0^{\infty} x^\lambda \phi(x)dx &=& \int_0^{1} x^\lambda
\left[\phi(x)-\phi(0)
-x\phi'(0) - \ldots - \frac{x^{n-1}}{(n-1)!} \, \phi^{(n-1)}(0)\right]dx \nonumber\\
&+& \int_1^{\infty} x^\lambda \phi(x)dx +
\sum_{k=1}^n\frac{\phi^{(k-1)}(0)}{(k-1)!(\lambda +k)}
 \ ,
\end{eqnarray}
which holds for $\lambda \neq -1,-2,\ldots,-n$. The above formula
shows that $\la \phi|x^\lambda_+\r$ as a function of $\lambda \in
\mathbb{C}$ has simple poles at $\lambda = -1,-2,\ldots$, and the
corresponding residue at $\lambda = -k$ equals
$\phi^{(k-1)}(0)/(k-1)!$.

Using the same arguments one shows that the distribution
$x^\lambda_-$ may be extended to the region $\mbox{Re}\, \lambda >
-n-1$ {\em via}:
\begin{eqnarray}  \label{pole-2}
\int_{-\infty}^{0} x^\lambda \phi(x)dx &=&  \int_0^{\infty}
x^\lambda  \phi(-x)dx
 \nonumber \\ &=&
\int_1^{\infty} x^\lambda \left[\phi(-x)-\phi(0) +x\phi'(0) -
\ldots - \frac{(-1)^{n-1}x^{n-1}}{(n-1)!} \,
\phi^{(n-1)}(0)\right]dx \nonumber\\
&+& \int_1^{\infty} x^\lambda \phi(x)dx +
\sum_{k=1}^n\frac{(-1)^{k-1}\phi^{(k-1)}(0)}{(k-1)!(\lambda +k)}
 \ ,
\end{eqnarray}
which holds for $\lambda \neq -1,-2,\ldots,-n$.
 Hence, $\la \phi|x^\lambda_-\r$  has simple poles at
$\lambda = -1,-2,\ldots$, and the corresponding residue at
$\lambda = -k$  equals $(-1)^{k-1}\phi^{(k-1)}(0)/(k-1)!$.

The Fourier transforms of $x^\lambda_\pm$
\begin{eqnarray}
F[x^\lambda_\pm](k) = \frac{1}{\sqrt{2\pi}} \int
e^{ikx}x^\lambda_\pm \,  dx\ ,
\end{eqnarray}
 are given by the following formula \cite{Gelfand-S}
\begin{eqnarray}  \label{F-x} F[x^\lambda_\pm](k) = \pm \frac{i}{\sqrt{2\pi}}
e^{\pm i\lambda\pi/2} \Gamma(\lambda + 1) (k+i0)^{-\lambda-1}\ ,
\end{eqnarray}
where $(k\pm i0)^\alpha$ is a distribution defined by:
\begin{eqnarray} (k \pm i0)^\alpha = k^\alpha_+ + e^{\pm i\alpha
\pi} k^\alpha_-\ .
\end{eqnarray} Due to the Euler
$\Gamma$-function the formula (\ref{F-x}) has single  poles at
$\lambda = -1,-2,\ldots$. Note, that although both $k^\alpha_+$
and $k^\alpha_-$ have poles at $\alpha = -1,-2,\ldots$, the
distribution $(k\pm i0)^\alpha$ is well defined for all $\alpha
\in \bbC$. Indeed
 \begin{eqnarray}
\lim_{\alpha \rightarrow -n} (k \pm i0)^\alpha = \lim_{\alpha
\rightarrow -n}
 (k^\alpha_+ + (-1)^n k^\alpha_-)\ ,
\end{eqnarray} and, due to (\ref{pole-1}) and (\ref{pole-2}),  the singular
parts of  $k^\alpha_+$ and $k^\alpha_-$, at $\alpha=-n$, cancel
out. In particular, for $\lambda = n  \in \bbN$, one obtains (cf.
\cite{Gelfand-S})
\begin{eqnarray}   \label{F-x-n} F[x^n_\pm](k) =
\frac{1}{\sqrt{2\pi}} \Big[ (\pm i)^{n+1} n! k^{-n-1} + (\mp
i)^n\pi \delta^{(n)}(k) \Big] \ .
\end{eqnarray}

\section*{Acknowledgments}

I would like to  thank J\c{e}drzej \'Sniatycki for very
interesting  discussions and his warm  hospitality during my stay
in Calgary and
 Andrzej Kossakowski for introducing this problem to me and for many
interesting and stimulating discussions.  This work was partially
supported by the Polish State Committee for Scientific Research
(KBN) Grant no 2P03B01619.

\end{document}